\documentclass[aps,pre,reprint,longbibliography,superscriptaddress,nobalancelastpage]{revtex4-2}

\usepackage{amsmath,amssymb,amsthm}
\usepackage{graphicx}
\usepackage{xcolor}
\usepackage{hyperref}
\usepackage{enumitem}
\usepackage{stmaryrd}
\usepackage{newtxtext,newtxmath}
\usepackage{accents}
\usepackage[compact]{titlesec}
\usepackage[formats]{listings}

\lstdefineformat{R}{~=\( \sim \)}
\definecolor{linkcolor}{HTML}{0000ff}
\hypersetup{colorlinks,linkcolor={linkcolor},citecolor={linkcolor},urlcolor={linkcolor}}

\renewcommand{\leq}{\leqslant}
\renewcommand{\geq}{\geqslant}

\newcommand{\figref}[2]{[Fig.~\hyperref[#1]{\ref*{#1}(#2)}]}
\newcommand{\figrefi}[2]{[Fig.~\hyperref[#1]{\ref*{#1}(#2)}, inset]}
\newcommand{\textfigref}[2]{Fig.~\hyperref[#1]{\ref*{#1}(#2)}}
\newcommand{\wholefigref}[1]{(Fig.~\ref{#1})}

\newcommand{\textwholefigref}[1]{Fig.~\ref{#1}}

\newcommand{\figrefp}[2]{\hyperref[#1]{\ref*{#1}(#2)}}
\newcommand{\citemm}[1]{\hyperref[#1]{Materials \& Methods, Sec.~\ref*{#1}}}
\newcommand{\textfigsuppref}[2]{\hyperref[#1]{Supplementary Fig.~\ref*{#1}(#2)}}
\newcommand{\wholefigsuppref}[1]{(\hyperref[#1]{Supplementary Fig.~\ref*{#1}})}
\newcommand{\textwholefigsuppref}[1]{\hyperref[#1]{Supplementary Fig.~\ref*{#1}}}

\DeclareMathAlphabet{\mathcal}{OMS}{cmsy}{m}{n}
\DeclareMathAlphabet{\mathbfsf}{\encodingdefault}{\sfdefault}{b}{n}
\DeclareMathAlphabet{\mathbfsfit}{\encodingdefault}{\sfdefault}{b}{it}
\renewcommand{\vec}[1]{\boldsymbol{#1}}
\newcommand{\tens}[1]{\mathbfsfit{#1}}
\renewcommand{\figurename}{Fig.}

\newtheorem*{theorem*}{Theorem}

\newcommand{\CC}{C\nolinebreak\hspace{-.05em}\raisebox{.4ex}{\tiny\bf+}\nolinebreak\hspace{-.10em}\raisebox{.4ex}{\tiny\bf +}}

\begin{document}
\title{Impossible ecologies: Interaction networks and stability of coexistence in ecological communities}
\author{Yu Meng}
\affiliation{Max Planck Institute for the Physics of Complex Systems, N\"othnitzer Stra\ss e 38, 01187 Dresden, Germany}
\affiliation{Center for Systems Biology Dresden, Pfotenhauerstra\ss e 108, 01307 Dresden, Germany}
\author{Szabolcs Horv\'{a}t}
\altaffiliation{Current address: Department of Computer Science, Reykjavík University, Menntavegur 1, 102 Reykjavík, Iceland.}
\affiliation{Center for Systems Biology Dresden, Pfotenhauerstra\ss e 108, 01307 Dresden, Germany}
\affiliation{\smash{Max Planck Institute of Molecular Cell Biology and Genetics, Pfotenhauerstra\ss e 108, 01307 Dresden, Germany}}
\author{Carl D. Modes}
\altaffiliation{Corresponding authors: modes@mpi-cbg.de, haas@pks.mpg.de.}
\affiliation{Center for Systems Biology Dresden, Pfotenhauerstra\ss e 108, 01307 Dresden, Germany}
\affiliation{\smash{Max Planck Institute of Molecular Cell Biology and Genetics, Pfotenhauerstra\ss e 108, 01307 Dresden, Germany}}
\affiliation{Cluster of Excellence Physics of Life, TU Dresden, Arnoldstra\ss e 18, 01069 Dresden, Germany}
\author{Pierre A. Haas}
\altaffiliation{Corresponding authors: modes@mpi-cbg.de, haas@pks.mpg.de.}
\affiliation{Max Planck Institute for the Physics of Complex Systems, N\"othnitzer Stra\ss e 38, 01187 Dresden, Germany}
\affiliation{Center for Systems Biology Dresden, Pfotenhauerstra\ss e 108, 01307 Dresden, Germany}
\affiliation{\smash{Max Planck Institute of Molecular Cell Biology and Genetics, Pfotenhauerstra\ss e 108, 01307 Dresden, Germany}}
\date{\today}
\begin{abstract}
Does an ecological community allow stable coexistence? Identifying the general principles that determine the answer to this question is a central problem of theoretical ecology. Random matrix theory approaches have uncovered the general trends of the effect of competitive, mutualistic, and predator-prey interactions between species on stability of coexistence. However, an ecological community is determined not only by the counts of these different interaction types, but also by their network arrangement. This cannot be accounted for in a direct statistical description that would enable random matrix theory approaches. Here, we therefore develop a different approach, of exhaustive analysis of small ecological communities, to show that this arrangement of interactions can influence stability of coexistence more than these general trends. We analyse all interaction networks of $N\leq 5$ species with Lotka--Volterra dynamics by combining exact results for $N\leq 3$ species and numerical exploration. Surprisingly, we find that a very small subset of these networks are ``impossible ecologies'', in which stable coexistence is non-trivially impossible. We prove that the possibility of stable coexistence in general ecologies is determined by similarly rare ``irreducible ecologies''. By random sampling of interaction strengths, we then show that the probability of stable coexistence varies over many orders of magnitude even in ecologies that differ only in the network arrangement of identical ecological interactions. Finally, we demonstrate that our approach can reveal the effect of evolutionary or environmental perturbations of the interaction network. Overall, this work reveals the importance of the full structure of the network of interactions for stability of coexistence in ecological communities.
\end{abstract}
\maketitle
Nigh on a century ago, Lotka and Volterra analysed a minimal model of two-species predator-prey dynamics~\cite{bacaer}. With extensions to competitive or mutualistic interactions and to interactions among $N\geq 2$ species together with multitudinous further generalisations, this became the generic description of an ecological community now known as the Lotka--Volterra model~\cite{* [] [{ Chap. 3 and Appendices, pp. 79--118 and 501--511.}] murray,hofbauer}. It is within such models that theoretical ecology can answer the paradigmatic question of stability: What are the general principles that determine the possibility of stable ecological coexistence? 

While the analysis of two-species incarnations of the Lotka--Volterra model is textbook material~\cite{murray}, not much analytical progress is possible for larger communities. Most understanding of these general principles therefore relies on the approach pioneered by May~\cite{may72}, who linked the statistics of stable coexistence to the eigenvalue distribution of a random matrix. Over the ensuing decades, as reviewed in Ref.~\cite{allesina15}, a large body of work explored how actual ecological communities overcome the stability limitations of May's simple random model~\cite{may72}. Meanwhile, on the mathematical side, further ideas from random matrix theory were adopted into theoretical ecology~\cite{allesina15}, although the link between random matrix theory \emph{à la} May and the Lotka--Volterra model was made explicit only recently~\cite{gibbs18}. In particular, this approach revealed generic effects of competitive, mutualistic, and predator-prey interactions in an ecological community on stability of coexistence~\cite{allesina12,mougi12,coyte15}: For example, increasing the proportion of predator-prey interactions stabilises coexistence, while increasing that of competitive or mutualistic interactions is destabilising~\cite{allesina12}, with mutualism being more destabilising than competition~\cite{allesina12,coyte15}.

The network of competitive, mutualistic, and predator-prey interactions in the community depends not only, however, on these proportions of different interaction types, but also on the network arrangement of these interactions. Recent work has explored the impact of community modularity~\cite{grilli16}, resource structures~\cite{butler18,butler20,gibbs22} and subpopulation structures associated with phenotypic switching~\cite{maynard19,haas20,haas22} on stability of coexistence, but this network structure, a more fundamental aspect of the structure of an ecological community, has garnered less attention. This is perhaps in part because it does not allow the direct statistical description in large ecological communities that allows random matrix theory approaches.

\begin{figure*}
\centering\includegraphics{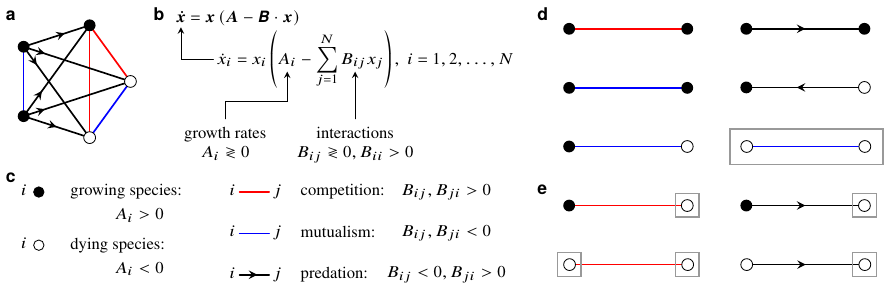}
\caption{Lotka--Volterra ecological dynamics and impossible ecologies. (a) Example of an ecological topology on $N=5$ species, defined below. (b) Mathematical definition of the Lotka--Volterra model on $N$ species: the dynamics of the vector $\vec{x}$ of the population abundances of the $N$ species are determined by a vector $\vec{A}$ of growth rates and a matrix $\tens{B}$ of interactions strengths. (c)~The \emph{ecological topology} is defined by the signs of $A_i\gtrless0$ and $B_{ij}\gtrless 0$ for $j\neq i$, defining competitive, mutualistic and directed predator-prey interactions between species. (d)~List of the six non-trivial ecological topologies on $N=2$ species: one topology (``obligate mutualism'', highlighted) is an \emph{impossible ecology} in which stable and feasible coexistence is non-trivially impossible. (e)~List of the four trivial ecologies on $N=2$ species: the highlighted species are dying on their own and have only deleterious interactions, prohibiting stable and feasible coexistence of all species.}\label{fig1}
\end{figure*}

Here, we reveal the huge effect of this network structure on stability of coexistence by taking a different approach, based on exhaustive analysis of small ecological communities: We analyse all interaction networks of $N\leq 5$ species with Lotka--Volterra dynamics. Combining exact calculations for $N\leq 3$ species and numerical exploration, we discover that stable coexistence is non-trivially impossible in a very small subset of these networks which we term ``impossible ecologies''. Somewhat conversely, we prove that any non-trivial network that contains a possible subecology is itself possible. The possibility of stable coexistence is therefore determined in general by the set of ``irreducible ecologies'' that are possible, but do not contain a possible subecology. Strikingly, our results suggest that these constitute an exponentially small fraction of all ecologies. Finally, we compute the probability of stable coexistence for all ecologies of $N\leq 5$ species by random sampling of interaction strengths. Remarkably, these probabilities vary, even for these small ecologies, over many orders of magnitude. This is true even in ecologies that have the same counts of interaction types, and hence differ only in the network structure of an identical set of ecological interactions. In this way, our results shift the stability paradigm partly from general stability principles to the detailed structure of the interaction networks.

\section*{\uppercase{Results}}
To understand the effect of the structure of interaction networks on ecological stability, we consider the simplest ecological interactions between $N$ species~\figref{fig1}{a} with Lotka--Volterra dynamics~\figref{fig1}{b}: a vector $\vec{x}=(x_1,x_2,\dots,x_N)$ of population abundances satisfies
\begin{subequations}\label{eq:lv}
\begin{align}
\vec{\dot{x}}=\vec{x}\left(\vec{A}-\tens{B}\cdot\vec{x}\right),  \label{eq:lv1}  
\end{align}
where $\vec{A}$ is a vector of growth rates and $\tens{B}$ is a matrix of interaction strengths. In components, this becomes
\begin{align}
\dot{x}_i=x_i\left(A_i-\sum_{j=1}^N{B_{ij}x_j}\right),\quad\text{for }i=1,2,\dots,N,    
\end{align}
\end{subequations}
where $A_i>0$ for a species $i$ growing on its own, $A_i<0$ for a species dying in the absence of inter-species interactions~\figref{fig1}{c}, and $B_{ii}>0$ defines the within-species competition of species~$i$. The interactions between species are defined by $B_{ij}\gtrless0$ for $j\neq i$ and may be of different types~\figref{fig1}{c}: the interaction between species $i$ and $j$ is competitive if $B_{ij},B_{ji}>0$; it is mutualistic if $B_{ij},B_{ji}<0$, and it is a directed predation interaction, with $i$ predating on $j$, if $B_{ij}<0$, $B_{ji}>0$. The signs of $A_i$ and $B_{ij}$ for $i,j=1,2,\dots,N$ and $j\neq i$ thus define the \emph{ecological topology}. 

Equations~\eqref{eq:lv} have a single equilibrium of coexistence of all species, $\vec{x_\ast}=\tens{B}^{-1}\cdot\vec{A}$, which we term \emph{feasible} if all population abundances at equilibrium are non-negative, i.e. $\min{\vec{x_\ast}}\geq 0$, and \emph{stable} if the dynamics return to $\vec{x_\ast}$ upon an infinitesimal perturbation away from it, which is if and only if all eigenvalues of the Jacobian matrix $-\vec{x_\ast}\tens{B}$ have negative real parts.

We therefore asked: how is the possibility of stable and feasible coexistence affected by ecological topology? To address this question statistically, we fixed an ecological topology, sampled its parameters, i.e., the magnitudes of the growth rates and interactions strengths, independently and uniformly from a uniform distribution, $|A_i|,|B_{ij}|\,\smash{\stackrel{\text{iid}}{\sim}}\,\mathcal{U}[0,1]$, and computed the probability $\mathbb{P}$ of $\vec{x_\ast}$ being stable and feasible.

The choice of Lotka--Volterra dynamics may appear to restrict our analysis, but, importantly, any dynamics reduce to the Lotka--Volterra dynamics~\eqref{eq:lv} in the vicinity of an equilibrium~(\citemm{methode}). Our analysis thus covers general ecological dynamics close to coexistence.

\subsection*{Two-species ecologies: impossible ecologies}
We first deployed our framework on the simplest case, of $N=2$ species. There are ten different ecological topologies on two species~\figref{fig1}{d),(e}. Each of the four topologies in \textfigref{fig1}{e} has (at least) one species, highlighted in \textfigref{fig1}{e}, that is dying on its own, and has only deleterious interactions with other species. The abundance of this species must therefore decrease, so there cannot be stable and feasible coexistence of all species in these topologies. We call such ecological topologies \emph{trivial}, and call other topologies, in which each species is growing on its own or has a favourable interaction with another species, \emph{non-trivial}. The remaining six ecologies on $N=2$ species~\figref{fig1}{d} are thus non-trivial. 

Interestingly, stable and feasible coexistence is only possible in five of them: In the final topology [``obligate mutualism'' highlighted in \textfigref{fig1}{d}], stable and feasible coexistence is non-trivially impossible, as proved in~\citemm{methoda}. We term such a topology an \emph{impossible ecology}.

\begin{figure*}
\centering\includegraphics{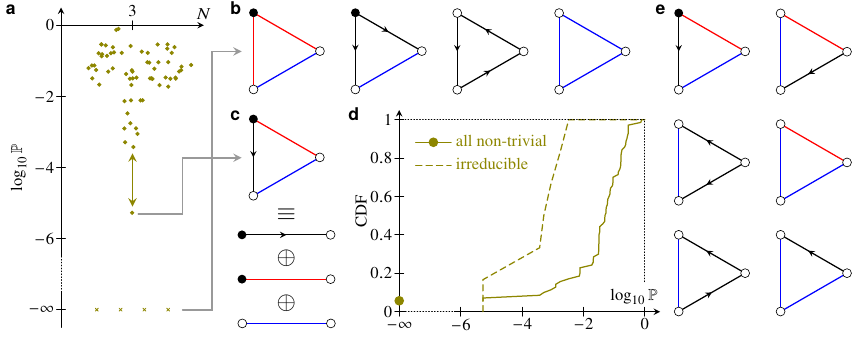}\vspace{-5pt}
\caption{Three-species ecologies: irreducible ecologies. (a) Swarm plot of the probability $\mathbb{P}$ of stable and feasible coexistence for all 70~non-trivial three-species ecologies. Impossible ecologies have $\log_{10}{\mathbb{P}}=-\infty$. (b) List of the four impossible ecologies on three species. (c)~Example of an irreducible ecology, decomposed into its three trivial or impossible two-species subecologies. (d) Cumulative distribution function of $\mathbb{P}$ for all 70 non-trivial three-species ecologies and restricted to the six irreducible ecologies. (e) List of the six irreducible ecologies on three species.
}\label{fig2}\vspace{-5pt}
\end{figure*}

\subsection*{Three-species ecologies: irreducible ecologies}
Next, we considered the case of $N=3$ species, for which there are 70 non-trivial ecologies~\wholefigsuppref{figS1}. Sampling magnitudes of growth rates and interaction strengths as described above, we computed numerical estimates of the probability $\mathbb{P}$ of stable and feasible coexistence for each of these topologies~\figref{fig2}{a}. These calculations suggested that $\mathbb{P}$ vanishes for four ecologies~\figref{fig2}{b}: these ecologies are indeed impossible ecologies, as proved in~\citemm{methodb}. These are competition of a growing species with two obligate mutualists, facultative predation of a growing species on two obligate mutualists, obligate cyclic predation, and obligate mutualism of three species. 

Among the remaining 66 possible ecologies, one has a value of $\mathbb{P}$ that is lower by orders of magnitude~\figref{fig2}{c}. Interestingly, the two-species subecologies of this ecology~\figref{fig2}{c} are either trivial or impossible. We therefore term this ecology an \emph{irreducible ecology}: It allows stable and feasible coexistence even though stable and feasible coexistence is possible in none of its subecologies. This definition suggests that irreducible ecologies might have low probabilities of stable and feasible coexistence. Indeed, the cumulative probability density function~\figref{fig2}{d} is shifted to lower probabilities upon restricting to the six irreducible ecologies on three species shown in \textfigref{fig2}{e}.

\subsection*{Extensions of ecologies}
To understand the importance of these irreducible ecologies, we proved a partial converse of these observations. This requires two definitions formalising our language: (1) We say that an ecological topology is \emph{possible} if it allows stable and feasible coexistence for some parameter values. (2) We say that an ecological topology is an \emph{extension} of another topology if the second can be obtained from the first by removing one species and its interactions with the remaining species. With these definitions, we have the following result:
\begin{theorem*}
Any non-trivial extension of a possible ecological topology is itself possible.
\end{theorem*}
Its proof is given in~\citemm{methodc}, broken down into two cases: in the first case, the added extension species grows on its own; in the second case, it does not.

\begin{figure}[b]
\vspace{-8pt}
\centering\includegraphics{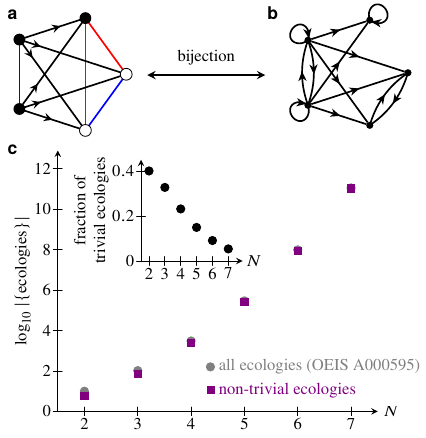}
\caption{Enumeration of ecologies. (a) Example of a five-species ecology as a complete graph with two-coloured nodes and four-coloured edges. (b) Image of this ecology under a bijection onto directed graphs with allowed self-loops. See text for definition of the bijection. (c) Plot of the number of all and non-trivial ecologies against the number $N$ of species, showing a combinatorial explosion. Inset: fraction of trivial ecologies plotted against $N$.
}\label{fig3}
\end{figure}

This general result has the strong consequence that the possibility of stable and feasible coexistence of any ecological topology is determined completely by the subset of irreducible ecological topologies. In other words, to understand the possibility of stable and feasible coexistence, it suffices to classify the irreducible ecologies. 

\begin{figure*}
\centering\includegraphics{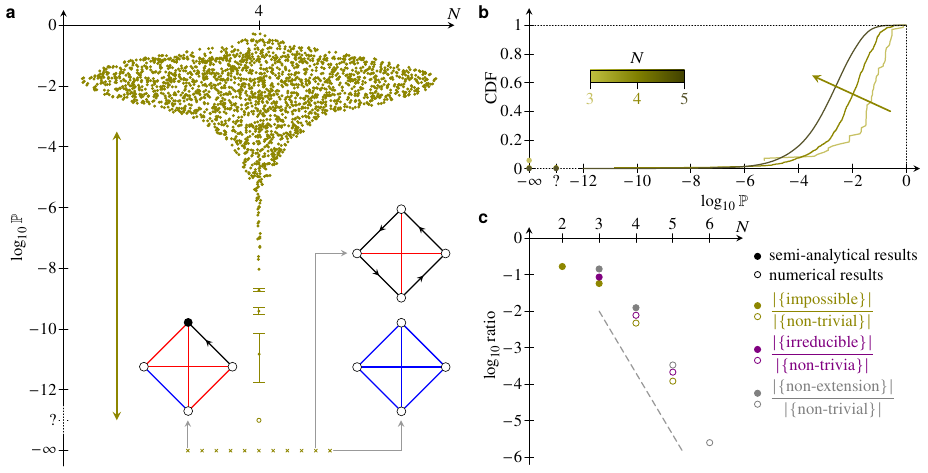}
\caption{Four- and five-species ecologies and the rareness of impossible and irreducible ecologies. (a)~Swarm plot of the probability $\mathbb{P}$ of stable and feasible coexistence for each of the 2340 non-trivial four-species ecologies. Identification of impossible ecologies ($\log_{10}{\mathbb{P}}=-\infty$) is based on numerical results only; we have no analytical proof of their impossibility. The question mark (?) indicates ecologies for which direct, non-uniform sampling of feasible equilibria yielded parameter values allowing stable and feasible coexistence, but for which no such parameter values could be found using uniform sampling and hence $\mathbb{P}$ could not be computed. Error bars are $95\%$ confidence intervals (\citemm{methodd}) larger than the plot markers. The vertical arrow emphasises that the probabilities vary over more than $12$ orders of magnitude. Insets: three examples of impossible ecologies; all impossible ecologies are listed in \textwholefigsuppref{figS2}. (b)~Cumulative distribution function of $\mathbb{P}$ for three-, four-, and five-species ecologies. The arrow emphasises the shift towards lower probabilities as the number~$N$ of species increases. (c)~Plot of the fraction of impossible, irreducible, and non-extension (i.e. irreducible or impossible) ecologies against~$N$ from semi-analytical and numerical computations. The dashed line suggests that these ratios decrease exponentially as $N$ increases.
}\label{fig4}\vspace{-5pt}
\end{figure*}

\subsection*{Larger ecologies}
Armed with this observation, we sought to extend our results to ecological topologies with $N>3$ species. First, we had to determine the non-trivial ecologies on $N$ species. This amounts to enumerating the graphs (more technically, the complete directed acyclic graphs) on $N$ two-coloured nodes (corresponding to positive or negative growth rates of the $N$ species) and with four-coloured edges (corresponding to the competitive, mutualistic, and two directed predator-prey interactions), as illustrated in~\textfigref{fig3}{a}. However, there is currently no direct, efficient algorithm for enumerating such node-and-edge-coloured graphs. We therefore transformed the problem by constructing a bijection between our ecological topologies and the set of directed graphs with allowed self-loops~\figref{fig3}{b}: A node will have a self-loop in the image graph under this mapping if it is growing, and no self-loop otherwise. Moreover, there are two edges between a pair of nodes in the image graph if the interaction between the corresponding species is mutualistic; there is no edge if it is competitive, and there is a single edge (directed towards the prey) if it is predation. By construction, an ecology is non-trivial if and only if each node in its image under this mapping has out-degree greater than zero. This mapping allows computationally efficient enumeration of both all and non-trivial ecologies using the \texttt{nauty} library~\cite{nauty}. Unsurprisingly, we found an extreme combinatorial explosion of the number of all and non-trivial ecologies alike~\figref{fig3}{c}. The former is a known sequence recorded in the On-line Encyclopedia of Integer Sequences (OEIS A000595) \cite{oeis}, while the latter, following the same trend, does not yet appear there. Interestingly, the fraction of ecologies that are trivial quickly decreases as the number of species increases~\figrefi{fig3}{c}.

Since the number of non-trivial ecologies reaches into the hundreds of millions and hundreds of billions for six and seven species, respectively, we cannot feasibly sample these ecologies sufficiently to calculate the associated probabilities for stable and feasible coexistence. However, four- and five-species ecologies \textit{are} within our grasp. Still, for these large-scale computations, we must use the \texttt{eigen} library for small matrix operations~\cite{eigen} liberally in our \CC{} code to make it sufficiently efficient (\citemm{methodd}).

\begin{figure*}
\centering\includegraphics{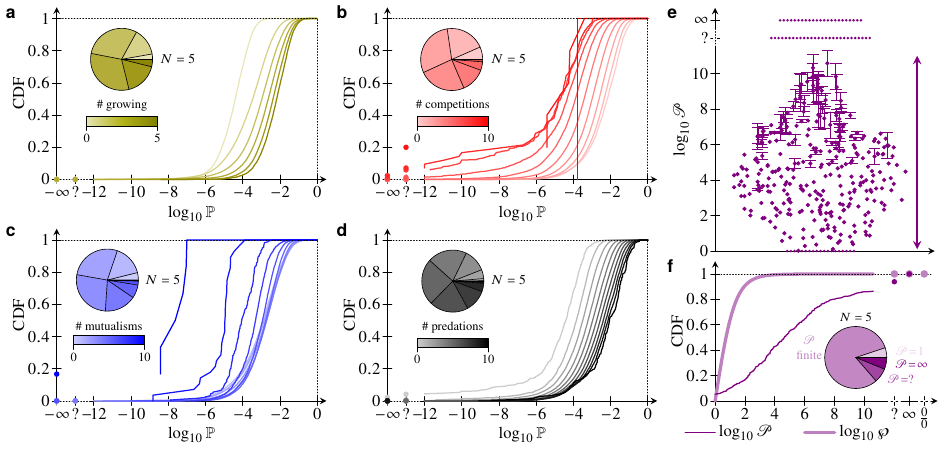}\vspace{-5pt}
\caption{Interplay of stability and ecological topology, analysed for all non-trivial ecologies with $N=5$ species. (a)~Cumulative distribution function (CDF) of the probability $\mathbb{P}$ of stable and feasible coexistence restricted to ecologies with $g$ species growing on their own, for $g=0,1,\dots,N=5$. $\log_{10}{\mathbb{P}}=-\infty$ corresponds to ecologies asserted to be impossible based on numerical calculations. The question~mark~(?) indicates ecologies for which direct, non-uniform sampling of feasible equilibria yielded parameter values allowing stable and feasible coexistence, but for which no such parameter values could be found using uniform sampling and hence $\mathbb{P}$ could not be computed. CDFs with more growing species are shifted to higher probabilities, showing that growing species are stabilising. The pie chart inset shows the distribution of the number of growing species in the different ecologies. (b)~Analogous plot of CDFs restricted to ecologies with $c$ competitive interactions between species, for $c=0,1,\dots,N(N-1)/2=10$, showing that competition is destabilising. (c)~Analogous plot of CDFs restricted to ecologies with $m$ mutualistic interactions between species, for $m=0,1,\dots,N(N-1)/2=10$, showing that high proportions of mutualistic interactions are destabilising. (d)~Analogous plot of CDFs restricted to ecologies with $p$ predator-prey interactions between species, for $p=0,1,\dots,N(N-1)/2=10$, showing that predation is stabilising. (e) Swarm plot, for $N=5$ species, of $\mathscr{P}$, defined in the text: This quantity measures the range of coexistence probabilities $\mathbb{P}$ for classes of ecologies that have the same number of growing species and the same number of mutualistic, competitive, and predator-prey interactions, so differ only ``purely topologically'' in the network arrangement of these interactions and growing species. The vertical arrow emphasises the many orders of magnitude over which $\mathscr{P}$ varies even for $N=5$ species. $\log_{10}{\mathscr{P}}=0$ corresponds to singleton classes; $\log_{10}{\mathscr{P}}=\infty$ corresponds to classes containing an impossible ecology. The question mark (?) indicates classes containing one of the ecologies for which $\mathbb{P}\not=0$ could not be computed. Error bars are $95\%$ confidence intervals (\citemm{methodd}) larger than the plot markers. (f) CDF of $\mathscr{P}$ and $\wp$, defined in the text, for the ecological classes with $N=5$ species. $\frac{0}{0}$ indicates undefined $\wp$. Inset: proportion of classes with unit, finite, unknown, and infinite range, respectively.
}\label{fig5}\vspace{-8pt}
\end{figure*}

\subsubsection*{Four-species ecologies}
We thus computed the probabilities of stable and feasible coexistence for all $2340$ non-trivial four-species ecologies. It is remarkable that they vary over more than $12$ orders of magnitude. In particular, we found $11$ impossible ecologies and $18$ irreducible ecologies (listed in \textwholefigsuppref{figS2}). We emphasise that the status of these $11$ impossible ecologies rests only on our numerical observations; we do not have analytical proofs of their impossibility to match those for two- and three-species ecologies. This number of $11$ impossible ecologies therefore represents an upper bound on the number of impossible ecologies (while the number of $18$ irreducible ecologies represents a lower bound). To provide further numerical support for our asserting these ecologies to be impossible, we also sampled feasible equilibria directly but not uniformly~(\citemm{methodd}), still without finding parameter instances allowing stable and feasible coexistence.

Three examples of ecologies that we assert to be impossible are shown as insets in \textfigref{fig4}{a}. One of these may be surprising as it is quite asymmetric. The other two are more symmetrical, and exhibit structures that we also found in three-species impossible ecologies, such as obligate mutualism or a cycle of obligate predation. We conjecture that these examples extend to infinite families of impossible ecologies, but do not have an analytical proof of this.

\subsubsection*{Five-species ecologies and the rareness of impossible and irreducible ecologies}
Similarly, we computed the probabilities of stable and feasible coexistence for all $248\,436$ non-trivial ecologies on five species. The cumulative distribution function of these probabilities is shifted towards lower probabilities compared to that for three- and four-species ecologies~\figref{fig4}{b}. 

We call \emph{non-extension ecologies} the non-trivial ecologies that do not contain a smaller possible subecology, so are impossible or irreducible by our definitions and our theorem. It follows that the non-extension ecologies on $N$ species are precisely the ecologies of which all the subecologies on $N-1$ species are trivial or impossible. With this observation, our list of impossible four-species ecologies yields an upper bound of 83 non-extension ecologies on $N=5$ species; these are listed in \textwholefigsuppref{figS3}. On sampling feasible equilibria directly again to identify possible and hence irreducible ecologies in this list (\citemm{methodd}), this leads to an upper bound of 30 impossible ecologies (again listed in \textwholefigsuppref{figS3}; this contains the five-species realisations of the infinite families conjectured to be impossible above). In turn, this list of impossible five-species ecologies species yields an upper bound of only $220$ non-extension topologies among the $88\,124\,470$ non-trivial ecologies of $N=6$ species. 

Importantly, these numbers represent tiny fractions of the total number of non-trivial ecologies and our numerical results actually suggest that these fractions decay exponentially as the number of species increases~\figref{fig4}{c}. The possibility of stable and feasible coexistence is thus completely determined by an exponentially small fraction of all non-trivial ecologies!

\subsection*{Interplay of stability and ecological topology}
We next focused more specifically on the interplay between stability and feasibility of coexistence and ecological topology by analysing our results for ecologies with $N=5$ species in more detail. First, we separated the cumulative distribution function (CDF) of $\mathbb{P}$~\figref{fig4}{b} according to the number of growing species in each ecological topology. The results in \textfigref{fig5}{a} show that the resulting CDFs are shifted towards higher probabilities the more growing species there are. In other words, species growth tends to stabilise ecologies. We performed a similar analysis for the number of competitive interactions, mutualistic interactions, and predator-prey interactions~\figref{fig5}{b)--(d}, which showed that competition and mutualism tend to be destabilising, while predation tends to be stabilising. These observations recover results from random matrix theory for large ecological communities~\cite{allesina12}. Additionally, our results suggest that mutualism is only destabilising if a large proportion of interactions are mutualistic: Indeed, the CDFs for small numbers of mutualistic interactions almost collapse on top of each other; only those for ecologies with a majority of mutualistic interactions are shifted towards low probabilities~\figref{fig5}{c}. More importantly, the distributions in \textfigref{fig5}{a)--(d} are wide, comparably so to the shift between the different distributions. This stresses the importance of the details of the interaction structure relative to the general stabilising or destabilising trends of species growth and interaction types.

To make this statement more quantitative, we noted that each ecological topology $e$ on $N=5$ species belongs to one and only
one class of ecologies $\mathscr{E}(g,c,m)$ with precisely $g$ growing species, $c$ competitive interactions, and $m$ mutualistic interactions, and hence with $p=N(N-1)/2-c-m$ predator-prey interactions. Ecologies belonging to the same class therefore have the same number of growing species and the same number of interactions of each type, so differ only in the network arrangement of these growing species and interactions. We define
\begin{subequations}
\begin{align}
\wp(g,c,m)=\left\{\max{\left\{\dfrac{\mathbb{P}(e)}{\mathbb{P}(e')},\dfrac{\mathbb{P}(e')}{\mathbb{P}(e)}\right\}}\,\biggr|\,e\not=e'\in\mathscr{E}(g,c,m)\right\},
\end{align}
the set of changes of probabilities of stable and feasible coexistence due to network rearrangements of an identical set of species and interactions between them. [We notice that $\wp(g,c,m)$ contains undefined elements if $\mathscr{E}(g,c,m)$ has more than one impossible ecology.] We also let
\begin{align}
\mathscr{P}(g,c,m)=\max{\wp(g,c,m)}=\dfrac{\displaystyle\max_{e\in\mathscr{E}(g,c,m)}{\mathbb{P}(e)}}{\displaystyle\min_{e\in\mathscr{E}(g,c,m)}{\mathbb{P}(e)}},
\end{align}
\end{subequations}
which thus quantifies the range of the probabilities of stable and feasible coexistence in each of these classes. Some classes are singleton classes, for which $\mathscr{P}=1$ by definition, and classes containing an impossible ecology have $\mathscr{P}=\infty$~\figref{fig5}{e),(f}. Strikingly, among the remaining classes that contain possible ecologies only, there are classes with $\mathscr{P}>10^{10}$~\figref{fig5}{e),(f} and more than half have $\mathscr{P}>10^4$~\figref{fig5}{f}, i.e., a range of more than four orders of magnitude, larger than the amount by which the medians of the distributions in~\textfigref{fig5}{a)--(d} are shifted. This shows that even for these small ecologies with ``only'' $N=5$ species, the effect of structural details on stability of coexistence is humongous and can swamp the effect of the number of growing species or of interactions of different types. However, comparing the distributions of $\mathscr{P}$ and $\wp$~\figref{fig5}{f} shows that even within a class with a large range, not all network rearrangements lead to a large change in coexistence probability.

\subsection*{Minimal perturbations of ecological topologies}
\begin{figure}[b]
\centering\includegraphics{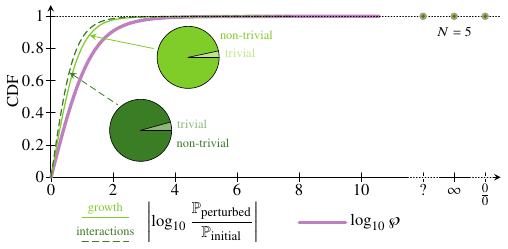}
\caption{Perturbations of ecological topologies. Cumulative distribution functions (CDFs) of the ratio of perturbed to initial probabilities of stable and feasible coexistence for all minimal topological perturbations of ecologies of $N=5$ species that change a single growing species into a dying one or vice versa (solid line, ``growth'') and all perturbations that change the competitive, mutualistic, or predator-prey nature of a single interaction between two species (dashed line, ``interactions''). CDFs are restricted to non-trivial perturbations; the insets show the proportion of trivial perturbations. Thick line: CDF of $\wp$, reproduced from \textfigref{fig5}{f} for comparison. The question mark~(?) indicates unknown values for perturbations involving ecological topologies for which $\mathbb{P}\not=0$ could not be computed; $\frac{0}{0}$ indicates perturbations for which the ratio of probabilities is undefined.}\label{fig6}
\end{figure}

To extend these results, we finally considered minimal perturbations of ecological topologies on $N=5$ species that might result from environmental or evolutionary pressures: For each ecological topology, we switched any single species from growing to dying (or vice versa) or we switched the nature of the competitive, mutualistic, or predator-prey interaction between any two species to a different one. For each such perturbation, changing an ecology $e$ into $e'$, say, we computed the fold-change of the probability of stable and feasible coexistence, $\left|\log_{10}{\mathbb{P}_{\text{perturbed}}/\mathbb{P}_{\text{initial}}}\right|=\left|\log_{10}{\mathbb{P}(e')/\mathbb{P}(e)}\right|$. (This ratio is infinite if exactly one of the initial and perturbed ecologies is impossible; it is undetermined if both are trivial, but the fraction of such perturbations is minute.) Interestingly, the fraction of perturbations that result in a trivial ecology is also small~(\textwholefigref{fig6}, insets). Importantly, the results reveal a certain robustness of coexistence to these perturbations compared to network rearrangements of the initial species and their interactions: The distribution of the changes of coexistence probability due to such perturbations is narrower than that of the changes due to such rearrangements~\wholefigref{fig6}. This average statement leaves open the question, of interest in the context of evolutionary or environmental pressures, whether specific perturbations are (more likely to be) stabilising or destabilising.

\section*{\uppercase{Discussion}}
In this paper, we have analysed the possibility of stable and feasible coexistence in all non-trivial interaction networks of $N\leq 5$ species. We found that a very small proportion of these are impossible ecologies, in which stable and feasible coexistence is non-trivially impossible. Somewhat conversely, we showed that the possibility of stable and feasible coexistence is completely determined by irreducible ecologies that constitute a similarly tiny proportion of all non-trivial ecologies. Finally, we revealed that, in spite of general trends of different interactions stabilising or destabilising coexistence, the details of the interaction network have a huge influence on the probability of stable and feasible coexistence.

In this way, our systematic, exhaustive analysis of ``small'' systems (still far beyond the reach of analytical approaches) complements random matrix theory \emph{à la} May~\cite{may72} by allowing us to address a problem that does not admit a direct statistical description in large ecological communities. Beyond theoretical ecology, in the rather different context of Turing instabilities of reaction-diffusion systems, this kind of approach has previously yielded insights into the robustness of Turing instabilities and the diffusive threshold inherent in the mechanism~\cite{scholes19,haas21}.  

Our results show that some ecological topologies, namely those that allow stable and feasible coexistence in large regions of parameter space, are much more robust to \emph{large} environmental fluctuations (i.e., large random parameter changes) than others that allow stable and feasible coexistence only in very small regions of parameter space. The question whether these topologies are robust to \emph{small} environmental fluctuations relates to the geometry of the subset of parameter space in which coexistence is stable and feasible and remains open. 

Moreover, our results concern coexistence at steady state. Extension to permanent coexistence beyond steady state, for example in a limit cycle or more complex attractor, requires going beyond Lotka--Volterra dynamics to analyse the effect of non-linearities including higher-order interactions, which are known to impact the stability of coexistence~\cite{grilli17,gibbs22b,gibbs23}, and constitutes another important challenge for the future. Future work will also need to understand how the details of the ecological topology affect another aspect of dynamics beyond steady state, namely the transients that can allow unstable ecological communities to persist over long timescales~\cite{hastings01,hastings18}.

Nevertheless, our results stress that the devil is in the details: the generic stabilising or destabilising trends of different interaction types~\cite{allesina12,mougi12,coyte15} can be swamped by the effect on stability and feasibility of coexistence of permutations of an identical set of competitive, mutualistic, and predator-prey interactions. Previous work~\cite{tang14} showed that, for permutation tests of interaction topology and intersection strength correlations of food webs with predator-prey relationships, the effect of these correlations on stability dominates over that of the interaction topology. However, the regions of parameter space in which coexistence is stable and feasible may be very different for different ecological topologies; this effect cannot be captured by such permutation tests, so this does not contradict our findings. In this context, the interaction network of competitive, mutualistic, and predator-prey interactions sets the (signs of) higher-order correlations between the entries of the interaction matrix. The fact that random matrix models including such higher-order correlations~\cite{korkmazhan21} display rich additional stability behaviour is thus consistent with our results.

Meanwhile, irreducible and impossible ecologies constrain crucial ecological processes: Irreducible ecologies restrict the paths of steady-state (dis)assembly of an ecological community towards stable and feasible coexistence~\cite{fukami11,angulo21,servan21,song21}. Impossible ecologies limit continued stable and feasible coexistence when the type of individual ecological interactions between species in a community changes due to environmental changes~\cite{callaway02,chamberlain14} or evolutionary pressures or when a species starts or stops to rely for its growth on resources, again due to environmental changes or evolutionary pressures. The rareness of these impossible and irreducible ecological topologies thus ascribes a certain robustness to these processes, already hinted at by our discussion of minimal ecological perturbations. This says that impossible and irreducible ecologies are more than mere mathematical curios: they have real ecological meaning.

Our results thus emphasise that generic stability principles can but paint an incomplete picture of the stability of coexistence in general ecological communities, which is inextricably linked to the detailed structure of the network of ecological interactions in the community. However, these principles of course still hold true in an average sense. A fascinating question is therefore: In real ecological communities, has evolution led to interaction networks that satisfy these principles, or has it selected those that break them?

\section*{\uppercase{Materials \& Methods}}
\subsection{Generality of the Lotka--Volterra equations near a coexistence equilibrium}\label{methode}
Here, we prove that general population dynamics of $N$ species linearise to Lotka--Volterra dynamics near a coexistence equilibrium; this result is probably folklore, but we could not find a reference. Consider general dynamics of a vector $\vec{x}=(x_1,x_2,\dots,x_N)$ of population abundances,
\begin{align}
\vec{\dot{x}}=\vec{f}(\vec{x}),\label{eq:gen}
\end{align}
or, in components, $\dot{x}_i=f_i(x_1,x_2,\dots,x_N)$ for $i=1,2,\dots,N$. We suppose that Eqs.~\eqref{eq:gen} have a coexistence equilibrium ${\vec{x_\ast}=(x_1^\ast,x_2^\ast,\dots,x_N^\ast)}$ such that $x_i^\ast\not=0$ for ${i=1,2,\dots,N}$. The Jacobian of Eqs.~\eqref{eq:gen} at $\vec{x_\ast}$ is $\tens{J}=(J_{ij})$. 

It now suffices to find $\vec{A}$ and $\tens{B}$ such that $\vec{x_\ast}$ is an equilibrium of the Lotka--Volterra equations~\eqref{eq:lv1}, and, by definition of linearisation, such that the Jacobians of Eqs.~\eqref{eq:lv1} and Eqs.~\eqref{eq:gen} at $\vec{x_\ast}$ are equal. The first condition is satisfied by imposing $\vec{A}=\tens{B}\cdot\vec{x_\ast}$, while the second requires $-\vec{x_\ast}\tens{B}=\tens{J}$. This can be satisfied by choosing $B_{ij}=J_{ij}/x_i^\ast$, which is well-defined by the above assumption on $\vec{x_\ast}$. This proves our claim.\hfill$\Box$
\subsection{Impossible ecologies of $\boldsymbol{N=2}$ species}\label{methoda}
Here, we prove that obligate mutualism of two species is an impossible ecology. The abundances of two such obligate mutualists obey
\begin{align}
&\dot{x}=x(-a-bx+cy), && \dot{y}=y(-d+ex-fy),
\end{align}
where $a,b,c,d,e,f$ are non-negative parameters. The coexistence equilibrium is
\begin{align}
&x_\ast=\frac{af+cd}{ce-bf}, &&y_\ast=\frac{ae+bd}{ce-bf}.
\end{align}
In particular, $x_\ast,y_\ast>0$ for feasibility, which yields $ce>bf$. The Jacobian at this equilibrium is
\begin{align}
\begin{pmatrix}
-bx_* & cx_*\\
ey_* & -fy_*
\end{pmatrix}.
\end{align}
Stability requires its determinant to be positive~\cite{murray}, which implies $(bf-ce)x_\ast y_\ast>0$. Since $x_\ast,y_\ast>0$ for feasibility, this yields $bf>ce$, which is a contradiction.\hfill$\Box$
\subsection{Impossible ecologies of $\boldsymbol{N=3}$ species}\label{methodb}
Here, we prove that obligate mutualism of three species, obligate cyclic predation of three species, facultative predation on two obligate mutualists, and competition with two obligate mutualists are impossible ecologies.
\subsubsection*{Obligate mutualism}
Consider obligate mutualism of three species, the abundances of which satisfy
\begin{subequations}\label{eq:m}
\begin{align}
\dot{x}&=x(-a-bx+cy+dz),\\
\dot{y}&=y(-e+fx-gy+hz),\\
\dot{z}&=z(-i+jx+ky-lz), 
\end{align}
\end{subequations}
wherein $a,b,c,d,e,f,g,h,i,j,k,l$ are non-negative parameters. The coexistence equilibrium $(x_\ast,y_\ast,z_\ast)$ has
\begin{subequations}\label{eq:meq}
\begin{align}
x_\ast&=\dfrac{k(ah-de)-(agl+cel+ich+dg)}{D},\\
y_\ast&=\dfrac{j(de-ah)-(ble+afl+ibh+idf)}{D},
\end{align}
\end{subequations}
where $D=bgl-bhk-cfl-chj-dfk-dgj$. Now assume feasibility. If $D>0$, then $x_\ast>0\Longrightarrow k(ah-de)>0$, but ${y_\ast>0\Longrightarrow j(de-ah)>0}$, so $ah\gtrless de$, a contradiction. Hence $D<0$. Now the characteristic polynomial of the Jacobian of Eqs.~\eqref{eq:m} at $(x_\ast,y_\ast,z_\ast)$ is $\lambda^3+a_1\lambda^2+a_2\lambda+a_3$, with $a_3=Dx_\ast y_\ast z_\ast$. One of the (necessary) Routh--Hurwitz conditions for stability is $a_3>0$~\cite{murray}. Since $D<0$, this implies $x_\ast y_\ast z_\ast<0$, contradicting feasibility.\hfill$\Box$
\subsubsection*{Obligate cyclic predation}
Consider three species with obligate cyclic predation, the abundances of which evolve according to
\begin{subequations}\label{eq:pp}
\begin{align}
\dot{x}&=x(-a-bx+cy-dz),\\
\dot{y}&=y(-e-fx-gy+hz),\\
\dot{z}&=z(-i+jx-ky-lz),
\end{align}
\end{subequations}
wherein $a,b,c,d,e,f,g,h,i,j,k,l$ are non-negative parameters. The coexistence equilibrium $(x_\ast,y_\ast,z_\ast)$ has
\begin{subequations}\label{eq:ppeq}
\begin{align}
x_\ast&=\dfrac{g(di-al)-(ahk+cel+chi+dek)}{D},\\
y_\ast&=\dfrac{f(al-di)-(ahj+bel+bhi+dej)}{D},
\end{align}
\end{subequations}
where $D=bgl+bhk+cfl-chj+dfk+dgj$. As in the case of obligate mutualism, $D<0$ because $D>0$ yields the contradiction $di\gtrless al$ from $x_\ast,y_\ast>0$. The Routh--Hurwitz conditions for stability then imply $x_\ast y_\ast z_\ast<0$ as in that case, contradicting feasibility again.\hfill$\Box$
\subsubsection*{Facultative predation on two obligate mutualists}
Consider facultative predation on two obligate mutualists, described by
\begin{subequations}\label{eq:ppm}
\begin{align}
\dot{x}&=x(a-bx+cy+dz),\\
\dot{y}&=y(-e-fx-gy+hz),\\
\dot{z}&=z(-i-jx+ky-lz),
\end{align}
\end{subequations}
wherein $a,b,c,d,e,f,g,h,i,j,k,l$ are non-negative parameters. The coexistence equilibrium $(x_\ast,y_\ast,z_\ast)$ has
\begin{subequations}\label{eq:ppmeq} 
\begin{align}
x_\ast&=\dfrac{g(al-di)-(ahk+cel+chi+dek)}{D},\\
y_\ast&=\dfrac{f(di-al)-(ahj+bel+bhi+dej)}{D},
\end{align}
\end{subequations}
where $D=dgj + chj + dfk - bhk + cfl + bgl$. As in the previous cases, $D<0$ because $D>0$ yields the contradiction $al\gtrless di$ from $x_\ast,y_\ast>0$. The Routh--Hurwitz conditions for stability then imply $x_\ast y_\ast z_\ast<0$ as in those cases, contradicting feasibility once again. \hfill$\Box$

It is remarkable that the seemingly worse case of \emph{obligate} predation on two obligate mutualists allows stable and feasible coexistence, so is an irreducible ecology!
\subsubsection*{Competition with two obligate mutualists}
Consider competition with two obligate mutualists, described by
\begin{subequations}\label{eq:ppc}
\begin{align}
\dot{x}&=x(a-bx-cy-dz),\\
\dot{y}&=y(-e-fx-gy+hz),\\
\dot{z}&=z(-i-jx+ky-lz),
\end{align}
\end{subequations}
wherein $a,b,c,d,e,f,g,h,i,j,k,l$ are non-negative parameters. The coexistence equilibrium $(x_\ast,y_\ast,z_\ast)$ has
\begin{subequations}\label{eq:ppceq} 
\begin{align}
y_\ast&=\dfrac{d(ej-fi)-(afl+ahj+bel+bhi)}{D},\\
z_\ast&=\dfrac{c(fi-ej)-(afk+agj+bek+bgi)}{D},
\end{align}
\end{subequations}
where $D=bgl-bhk-cfl-chj-dfk-dgj$. As in the previous cases, $D<0$ because $D>0$ yields the contradiction $ej\gtrless fi$ from $y_\ast,z_\ast>0$. The Routh--Hurwitz conditions for stability then imply $x_\ast y_\ast z_\ast<0$ as in those cases, contradicting feasibility once again.\hfill$\Box$ 
\subsection{Proof of the theorem on extensions of ecological topologies}\label{methodc}
Here, we prove the theorem on extensions of ecological topologies stated in the main text:
\begin{theorem*}
Any non-trivial extension of a possible ecological topology is itself possible.
\end{theorem*}
\emph{Proof.} Let $N\geq 2$ be an integer. Consider a possible topology of $N-1$ species, the abundances $\vec{x_0}$ of which follow
\begin{align*}
\vec{\dot{x}_0}=\vec{x_0}(\vec{A_0}-\tens{B}_{\mathbfsf{0}}\cdot\vec{x_0}),    
\end{align*}
with stable and feasible coexistence equilibrium $\vec{x_0^\ast}=\tens{B}_{\mathbfsf{0}}^{-1}\cdot\vec{A_0}$, and associated Jacobian $\tens{J}_{\mathbfsf{0}}=-\vec{x_0^\ast}\tens{B}_{\mathbfsf{0}}$ with stable eigenvalues $\mathcal{E}(\tens{J}_{\mathbfsf{0}})$. Choose $\varepsilon\ll 1$ such that
\begin{align}
\min{|\vec{A_0}|},\min{|\tens{B}_{\mathbfsf{0}}|},\min{\vec{x_0^\ast}},\min{|\text{Re}(\mathcal{E}(\tens{J_{\mathbfsf{0}}}))|}\gg\varepsilon.\tag{$\ddag$}     
\end{align}
We extend this ecology non-trivially by adding species $N$, with abundance $x_N$. This extended ecology is described by the block equation
\begin{align*}
\vec{\dot{x}}=\vec{x}\left[\left(\begin{array}{c}
\vec{A_0}\\\hline a
\end{array}\right)-\left(\begin{array}{c|c}
\tens{B}_{\mathbfsf{0}}&\vec{b}\\\hline\vec{c}^\top&d
\end{array}\right)\cdot\vec{x}\right],\quad\text{with }\vec{x}=\left(\begin{array}{c}
\vec{x_0}\\\hline x_N
\end{array}\right),        
\end{align*}
and where $a>0$ if species $N$ grows on its own, $a<0$ if species $N$ dies in the absence of inter-species interactions, and where $d>0$ defines its within-species competition, and the vectors $\vec{b},\vec{c}$ determine its interactions with the other species. We denote by $\vec{x_\ast}$ the corresponding coexistence equilibrium, and $\tens{J}$ the associated Jacobian. The proof now divides into two cases:
\begin{enumerate}[label=(\arabic{enumi}),leftmargin=*]
\item In the first case, $a>0$, and we choose $a,d=O(1)$ and $\vec{b},\vec{c}=O(\varepsilon)$, so that
\begin{align*}
\vec{\dot{x}}=\vec{x}\left[\left(\begin{array}{c}
\vec{A_0}\\\hline a
\end{array}\right)-\left(\begin{array}{c|c}
\tens{B}_{\mathbfsf{0}}&O(\varepsilon)\\\hline O(\varepsilon)&d
\end{array}\right)\cdot\vec{x}\right].
\end{align*}
We now use assumptions ($\ddag$) repeatedly to obtain, first,
\begin{align*}
\hphantom{AA}\vec{x_\ast}=\left(\begin{array}{c|c}
\tens{B}_{\mathbfsf{0}}&\vec{0}\\\hline \vec{0}^\top&d
\end{array}\right)^{-1}\cdot\left(\begin{array}{c}
\vec{A_0}\\\hline a
\end{array}\right)+O(\varepsilon)=\left(\begin{array}{c}
\vec{x_{\smash0}^\ast}\\\hline a/d
\end{array}\right)+O(\varepsilon),
\end{align*}
and thence
\begin{align*}
\hphantom{AA}\tens{J}=-\left(\begin{array}{c}
\vec{x_{\smash0}^\ast}\\\hline a/d
\end{array}\right)\left(\begin{array}{c|c}
\tens{B}_{\mathbfsf{0}}&\vec{0}\\\hline \vec{0}^\top&d
\end{array}\right)+O(\varepsilon)=\left(\begin{array}{c|c}
\tens{J}_{\mathbfsf{0}}&\vec{0}\\\hline \vec{0}^\top&-a
\end{array}\right)+O(\varepsilon).            
\end{align*}
Thus $\vec{x_\ast}$ is feasible. Moreover, $\mathcal{E}(\tens{J})=\{\mathcal{E}(\tens{J}_{\mathbfsf{0}}),-a\}+O(\varepsilon)$, whence $\vec{x_\ast}$ is stable, too. 
\item In the second case, $a>0$. Let $\vec{c}=(c_1,c_2,\dots,c_{N-1})$. Because the extension is non-trivial by assumption, there exists $i\in\{1,2,\dots,N-1\}$ such that $c_i<0$. Without loss of generality, $i=1$ and we write $\vec{c}=(c_1,\vec{c'})$. We now choose $a=O(\varepsilon)$, $\vec{b}=O(\varepsilon)$, $c_1=O(\varepsilon^{1/2})$, $\vec{c'}=O(\varepsilon)$, $d=O(1)$. Writing $c_1=-\varepsilon^{1/2}\gamma$, where $\gamma=O(1)$ and $\gamma>0$, we obtain
\begin{align*}
\vec{\dot{x}}=\vec{x}\left[\left(\begin{array}{c}
\vec{A_0}\\\hline O(\varepsilon)
\end{array}\right)-\left(\begin{array}{c|c|c}
\multicolumn{2}{c|}{\tens{B}_{\mathbfsf{0}}}&O(\varepsilon)\\\hline -\varepsilon^{1/2}\gamma&O(\varepsilon)&d
\end{array}\right)\cdot\vec{x}\right],
\end{align*}
yielding the equilibrium condition
\begin{align*}
\left(\begin{array}{c|c|c}
\multicolumn{2}{c|}{\tens{B}_{\mathbfsf{0}}}&\vec{0}^\top\\\hline -\varepsilon^{1/2}\gamma&\vec{0}&d
\end{array}\right)\cdot\vec{x_\ast}  =\left(\begin{array}{c}
\vec{A_0}\\\hline 0
\end{array}\right)+O(\varepsilon).  
\end{align*}
We decompose
\begin{align*}
\vec{x_\ast}=\left(\begin{array}{c}
\vec{\chi_\ast}\\\hline x_N^\ast
\end{array}\right)+O(\varepsilon).    
\end{align*}
The first components of the equilibrium condition become 
\begin{align*}
\tens{B}_{\mathbfsf{0}}\cdot\vec{\chi_\ast}=\vec{A_0}\quad\Longrightarrow\quad \vec{\chi_\ast}=\tens{B}_{\mathbfsf{0}}^{-1}\cdot\vec{A_0}=\vec{x_0^\ast}.   
\end{align*} 
Writing $\vec{x_0^\ast}=(\xi^\ast,\vec{\xi'_\ast})$ with in particular $\xi^\ast>0$, its final component then yields
\begin{align*}
-\varepsilon^{1/2}\gamma\xi^\ast+dx_N^\ast=0\quad\Longrightarrow\quad x_N^\ast=\varepsilon^{1/2}\dfrac{\gamma}{d}\xi^\ast.
\end{align*}
In particular, this shows that $\vec{x_\ast}$ is feasible. Moreover,
\begin{align*}
\tens{J}&=-\left(\begin{array}{c}
\vec{x_{\smash0}^\ast}\\\hline \varepsilon^{1/2}\gamma \xi^\ast/d
\end{array}\right)\left(\begin{array}{c|c|c}
\multicolumn{2}{c|}{\tens{B}_{\mathbfsf{0}}}&\vec{0}^\top\\\hline -\varepsilon^{1/2}\gamma&\vec{0}&d
\end{array}\right)+O(\varepsilon)\\
&=\left(\begin{array}{c|c}
\tens{J}_{\mathbfsf{0}}&\vec{0}\\\hline \vec{0}^\top&-\varepsilon^{1/2}\gamma \xi^\ast
\end{array}\right)+O(\varepsilon),
\end{align*}
which has eigenvalues $\mathcal{E}(\tens{J})=\{\mathcal{E}(\tens{J}_{\mathbfsf{0}}),-\varepsilon^{1/2}\gamma \xi^\ast\}+O(\varepsilon)$, which are stable.
\end{enumerate}
Stable and feasible coexistence is thus possible in the extended ecological topology in either case. This completes the proof of the theorem.\hfill$\Box$
\subsection{Numerical Methods}\label{methodd}
\subsubsection*{Numerical tolerances}
To avoid numerical errors in the computation of deciding whether coexistence is stable and feasible for randomly sampled parameter values $|\vec{A}|,|\tens{B}|$, we replace, for numerical purposes, the exact feasibility and stability conditions with $\min{\vec{x_\ast}}>\tau$ and $\max{\mathrm{Re}(-\vec{x_\ast}\tens{B})}<-\tau$, respectively, to which conditions we add $\left|\det{\tens{B}}\right|>\tau$. Our tests (not shown) show that, by choosing $\tau=10^{-7}$, this avoids errors due to the entries of $\vec{x_\ast}$ varying by more than $(\text{machine precision})^{-1}$.

Somewhat conversely, we declare systems with $\left|\min{\vec{x_\ast}}\right|\leq\tau$ or  $\left|\max{\mathrm{Re}(-\vec{x_\ast}\tens{B})}\right|\leq\tau$ or $\left|\det{\tens{B}}\right|\leq\tau$ to be uncertain. Thus, after $n$ samplings, of which $c$ allowed stable and feasible coexistence and $u$ were uncertain, we estimate the interval $[c/n,(c+u)/n]$ for the probability of stable and feasible coexistence. We add $95\%$ Wilson confidence intervals (see, e.g., Ref.~\cite{brown01}) to the endpoints of this interval to define the error bars reported in Figs.~\ref{fig4} and~\ref{fig5}. 
\subsubsection*{Direct sampling of feasible equilibria}
Feasible equilibria of a non-trivial ecology can be sampled directly (but non-uniformly in parameter space) by the following algorithm:
{\small\begin{verbatim}
input:  interaction_signs[N][N], growth_signs[N]
output: x[N], A[N], B[N,N]

for i = 1..N:
    x[i] = random
    for j = 1..N:
        B[i,j] = interaction_signs[i,j]*random
A = B*x                            # matrix product
for i such that growth_signs[i] == +:
    d = A[i] - B[i][i]*x[i]
    if d < 0:
        B[i][i] = (-d+random)/x[i]
for i such that growth_signs[i] == -:
    find j such that interaction_signs[i,j] == -
    d = A[i] - B[i][j]*x[j]
    if d > 0:
        B[i][j] = (-d-random)/x[j]
\end{verbatim}}
\noindent This algorithm generates a feasible equilibrium $\vec{x_\ast}$ and samples a random matrix of interaction strengths $\tens{B}$ consistent with the topology. It then adjusts one term in each row of $\tens{B}$, while keeping it consistent with the topology, to ensure that $\vec{A}=\tens{B}\cdot\vec{x_\ast}$ is consistent with the topology; we stress that the \texttt{find} instruction in the final loop must return a result because the topology is non-trivial by assumption.

This choice of algorithm is not of course unique, but we found that it allowed us to find parameter values allowing stable and
feasible coexistence for topologies for which we were unable to find such parameter values by uniform sampling of parameters.
\subsubsection*{Enumeration of (non-extension) ecologies}
By our theorem on extensions of ecological topologies, an ecology on $N$ species is a non-extension ecology if and only if all its non-trivial subecologies on $N-1$ species are impossible.

To classify non-extension ecologies on $N$ species using \texttt{nauty}~\cite{nauty} and the bijection discussed in the main text, we therefore first generate all non-trivial ecologies using
{\small\begin{lstlisting}
geng N | directg | vcolg -m2 -o | pickg -~d0
\end{lstlisting}}
\noindent in which the final pipe removes graphs that have vertices with zero out-degree, hence are trivial. We obtain all non-trivial subecologies by piping the result to the \texttt{nauty} command \texttt{delptg -m1}, of which we modified the output format to track which ecology each subecology in the resulting list arises from. We then generated adjacency matrices of all impossible ecologies on $N-1$ species and their permutations, and converted them to \texttt{digraph6} format using the \texttt{nauty} tool \texttt{amtog -z}, which we piped to \texttt{uniq} to remove repeats. Because of the small number of impossible ecologies, we could then find the the impossible subecologies on $N-1$ species and hence the non-extension ecologies on $N$ species by brute force.
\section*{}
\paragraph*{Author contributions.} YM, CDM, and PAH designed the study; YM, SzH, PAH wrote code and analysed data; SzH and PAH derived analytical results; YM, CDM, PAH analysed and interpreted results; YM and PAH wrote the initial draft of the paper; all authors contributed to the final draft of the paper.
\paragraph*{Acknowledgements.} The authors thank Alexander Wietek for suggesting the use of the \texttt{eigen} library~\cite{eigen} and Sebastian Eibl, Malcolm Hillebrand, Markus Rampp, and Hubert Scherrer-Paulus for discussions on \CC{} code efficiency, and gratefully acknowledge funding from the Max Planck Society.
\paragraph*{Data and code availability.} \CC{} sample code and data are available at [url to be inserted].

\bibliography{manuscript}

\newpage
\onecolumngrid
%\section*{Supplementary Figures}
\setcounter{figure}{0}
\renewcommand{\figurename}{Supplementary Fig.}
\begin{figure*}[h]
\centering\includegraphics{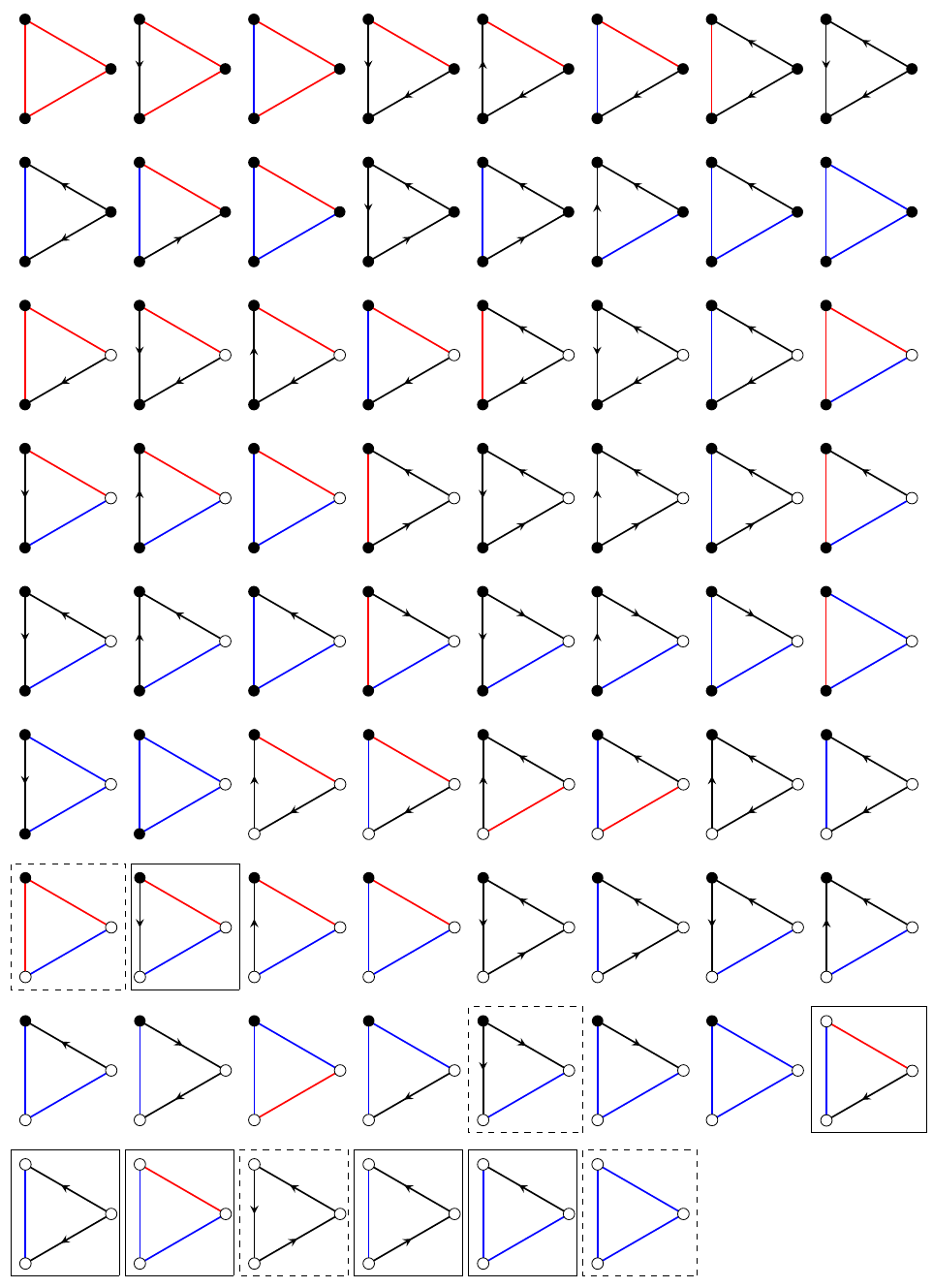}\vspace{-5pt}
\caption{List of all 70 non-trivial ecologies on three species. Impossible and irreducible ecologies are highlighted by dashed and solid boxes, respectively.}\label{figS1}
\end{figure*}
\begin{figure*}[h]
\centering\includegraphics{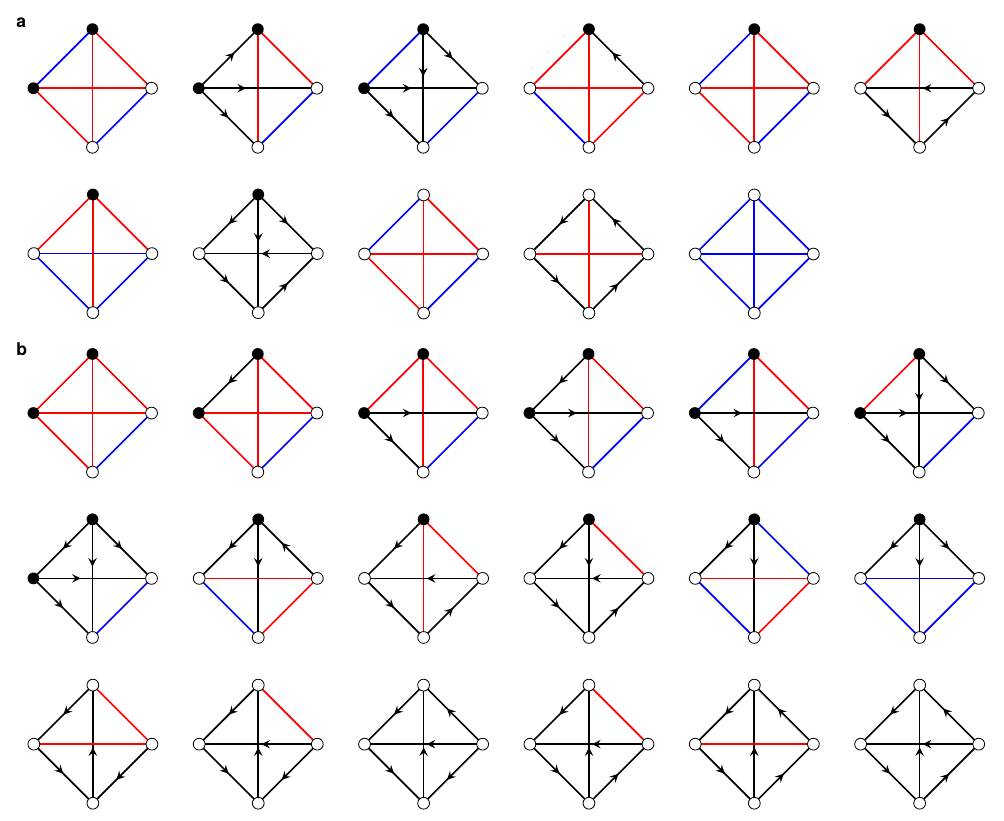}
\caption{List of the 29 non-extension ecologies that are found among the 2340 non-trivial ecologies of four species . (a) List of 11 ecologies asserted to be impossible; as discussed in the main text, this is an upper bound based on numerical calculations, and we do not have an analytical proof of impossibility for any of these ecologies. (b) List of the remaining 18 non-extension ecologies of four species, shown to be irreducible by direct non-uniform sampling of feasible equilibria yielding parameter values allowing stable and feasible coexistence.}\label{figS2}
\end{figure*}
\begin{figure*}[h]\vspace{-5pt}
\centering\includegraphics{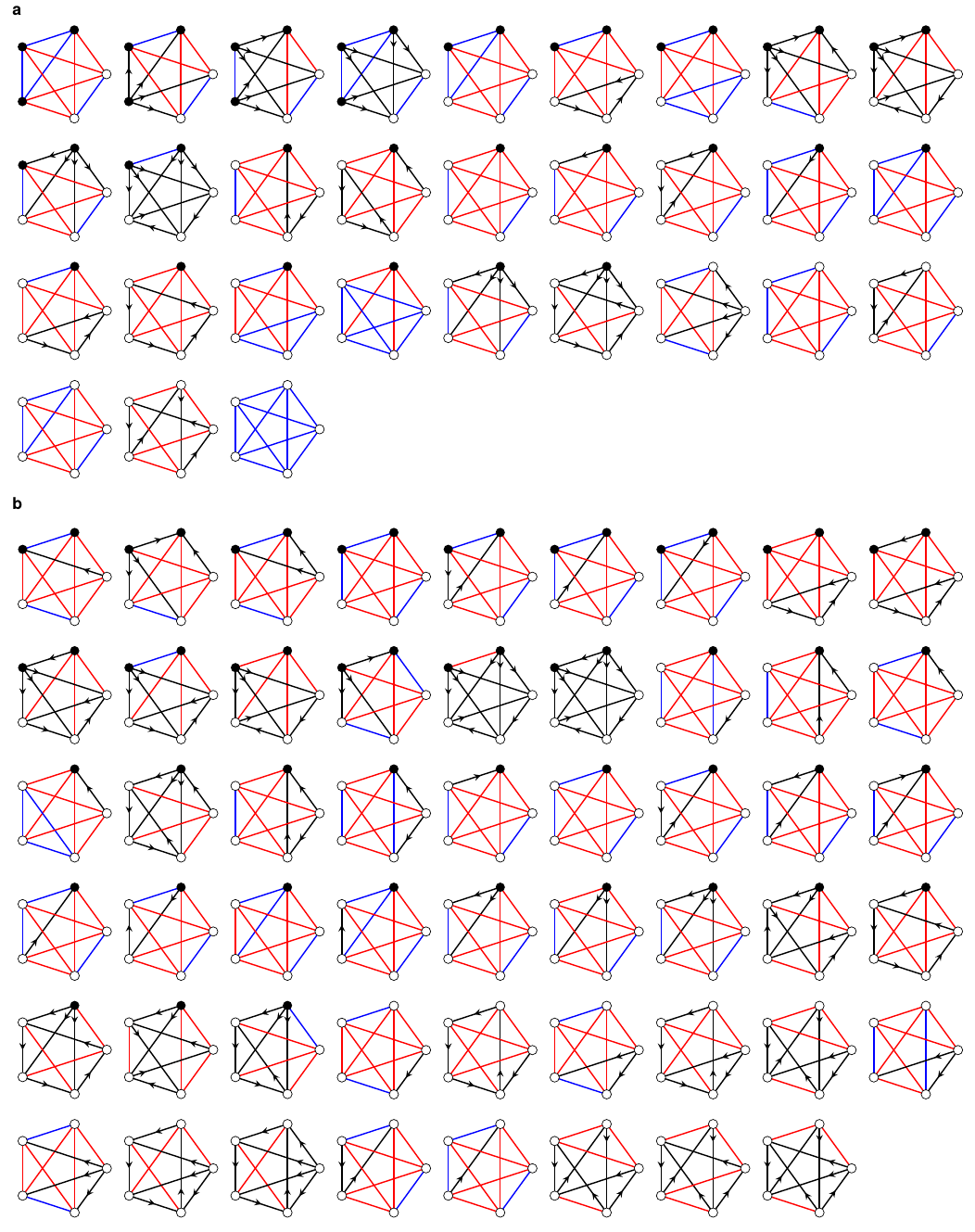}\vspace{-9pt}
\caption{List of the 83 non-extension ecologies that are found among the 248436 non-trivial ecologies of five species, assuming that the ecologies in \textfigsuppref{figS2}{a} are indeed impossible. (a) List of 30 ecologies asserted to be impossible, similarly to the four-species case. (b) List of the remaining non-extension ecologies, shown to be irreducible as in the four-species case.}\label{figS3}
\end{figure*}
\end{document}